\documentclass[reprint,superscriptaddress,amsmath,amssymb,aps,prl,
]{revtex4-1}

\usepackage{graphicx}
\usepackage{dcolumn}
\usepackage{bm}
\usepackage{xcolor}

\begin{document}

\title{Floquet engineering of black phosphorus upon below-gap pumping}

\author{Shaohua Zhou}
 \altaffiliation{These authors contributed equally to this work}
 \affiliation{State Key Laboratory of Low-Dimensional Quantum Physics and Department of Physics, Tsinghua University, Beijing 100084, China}

\author{Changhua Bao}
 \altaffiliation{These authors contributed equally to this work}
 \affiliation{State Key Laboratory of Low-Dimensional Quantum Physics and Department of Physics, Tsinghua University, Beijing 100084, China}

\author{Benshu Fan}
 \altaffiliation{These authors contributed equally to this work}
 \affiliation{State Key Laboratory of Low-Dimensional Quantum Physics and Department of Physics, Tsinghua University, Beijing 100084, China}

\author{Fei Wang}
 \affiliation{State Key Laboratory of Low-Dimensional Quantum Physics and Department of Physics, Tsinghua University, Beijing 100084, China}

\author{Haoyuan Zhong}
 \affiliation{State Key Laboratory of Low-Dimensional Quantum Physics and Department of Physics, Tsinghua University, Beijing 100084, China}

\author{Hongyun Zhang}
 \affiliation{State Key Laboratory of Low-Dimensional Quantum Physics and Department of Physics, Tsinghua University, Beijing 100084, China}

\author{Peizhe Tang}
 \altaffiliation{Corresponding author: peizhet@buaa.edu.cn}
 \affiliation{School of Materials Science and Engineering, Beihang University, Beijing 100191, China}
 \affiliation{Max Planck Institute for the Structure and Dynamics of Matter, Center for Free Electron Laser Science, 22761 Hamburg, Germany}

\author{Wenhui Duan}
 \affiliation{State Key Laboratory of Low-Dimensional Quantum Physics and Department of Physics, Tsinghua University, Beijing 100084, China}
 \affiliation{Frontier Science Center for Quantum Information, Beijing 100084, China}
 \affiliation{Institute for Advanced Study, Tsinghua University, Beijing 100084, China}

\author{Shuyun Zhou}
 \altaffiliation{Corresponding author: syzhou@mail.tsinghua.edu.cn}
 \affiliation{State Key Laboratory of Low-Dimensional Quantum Physics and Department of Physics, Tsinghua University, Beijing 100084, China}
 \affiliation{Frontier Science Center for Quantum Information, Beijing 100084, China}

\date{\today}

\begin{abstract}
Time-periodic light field can dress the electronic states and lead to light-induced emergent properties in quantum materials. While below-gap pumping is regarded favorable for Floquet engineering, so far direct experimental evidence of momentum-resolved band renormalization still remains missing. Here, we report experimental evidence of light-induced band renormalization in black phosphorus by pumping at photon energy of 160 meV which is far below the band gap, and the distinction between below-gap pumping and near-resonance pumping is revealed. Our work demonstrates light-induced band engineering upon below-gap pumping, and provides insights for extending Floquet engineering to more quantum materials. 
\end{abstract}

\maketitle


Light-matter interaction can be utilized not only for probing the intrinsic properties of quantum materials, but also for tailoring the non-equilibrium material properties by forming hybridized states between electrons and photons, so-called Floquet states \cite{Oka2009PhotoHE}. The formation of light-dressed Floquet states can result in the modification of the transient electronic structure of quantum materials, inducing fascinating emergent phenomena \cite{okaRev2019,Lindner2020,HsiehDemond2017,DelaTorre2021,BaoNatRevPhys,Devereaux2018NRP}, such as photo-Hall effect \cite{Oka2009PhotoHE,Demler2011PhotoHE,CavalleriNP20}, Floquet topological insulators \cite{Lindner2011natphy,Podolsky2013PRL,Devereaux2016NC}, transient topological phase transitions \cite{zhang2016theory,WangzhongPRL2016,Rubio2017nc}, and modification of nonlinear optical properties \cite{Hsieh2021nat} and excitons \cite{Ghimire2023np} etc.

The Floquet interaction strength scales inversely with $\hbar\omega$ \cite{Park2014Theory,Devereaux2018NRP}, suggesting that Floquet engineering can in principle be enhanced by reducing the pump photon energy $\hbar\omega$ (Fig.~1a, middle panel).  Moreover, since there is no direct optical absorption upon below-gap pumping, fewer dissipation channels are activated and therefore heating effects can be suppressed, which is also favorable for realizing Floquet engineering. 
Such below-gap pumping could induce a large band renormalization as schematically illustrated in Fig.~1b, resulting in phenomena such as optical Stark effect in conventional semiconductors \cite{Townes1955PR,Reimann1985prl} and valley-selective optical Stark effect in transition metal dichalcogenides \cite{Gedik2015Stark,Gediksci2017}. More recently, ultrafast modulation of the optical nonlinearity has been reported in MnPS$_3$ upon below-gap pumping \cite{Hsieh2021nat}, which is also related to the Floquet engineering.  Time- and angle-resolved photoemission spectroscopy (TrARPES) is a powerful technique for revealing the transient light-induced Floquet band engineering. So far, while Floquet band engineering has been directly visualized upon resonance pumping by TrARPES \cite{WangYH2013,GedikNatPhys2016,Zhou2023Nat,HuberNat2023}, momentum-resolved band renormalization upon below-gap pumping still remains missing. Experimental realization of Floquet engineering upon below-gap pumping is an important step to extend Floquet engineering to a wider range of quantum materials, because it does not require resonance conditions to match the pump photon energy with the gap value. 

\begin{figure}[htbp]
	\centering
	\includegraphics{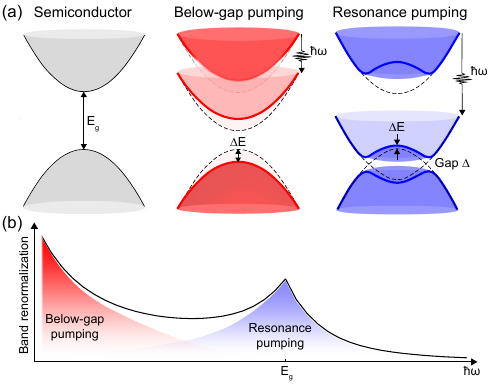}
	\caption{(a) Schematic dispersions near the band edge of black phosphorus in the equilibrium state (left panel), upon below-gap pumping (middle panel) and near-resonance pumping (right panel). (b) Schematic light-induced band renormalization at the $\Gamma$ point $\Delta E$ with pump photon energy. }
	\label{Fig1}
\end{figure}

Here, we report the experimental observation of Floquet band engineering of black phosphorus  upon below-gap pumping by TrARPES measurements. A momentum-dependent band renormalization is observed  near the valence band edge. Moreover, the evolution of the band renormalization from below-gap pumping  to near-resonance pumping is revealed from both experimental results and theoretical calculations, and their distinctions and implications are discussed.

\begin{figure}[htbp]
	\centering
	\includegraphics{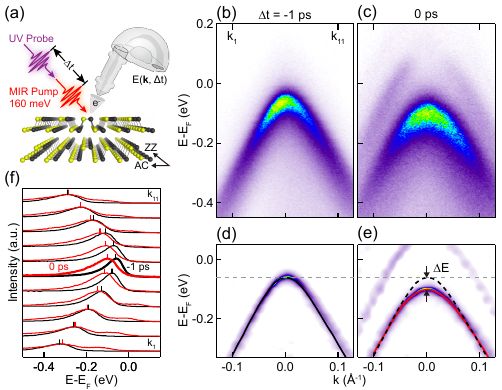}
	\caption{(a) A schematic of TrARPES with MIR pumping and the atomic structure of black phosphorus.  (b-e) TrARPES dispersion images measured along the AC direction (b,c) and corresponding second derivative images (d,e) by $s$-$pol.$ pump at $\Delta t = $-1 ps (b,d) and 0 ps (c,e) with pump photon energy of 160 meV and a pump fluence of 500 $\mu$J/cm$^2$.  (f) EDCs for data at $\Delta t = $-1 ps (black curves) and $\Delta t = $ 0 (red curves) at the momentum points marked in (b,c).}
	\label{Fig2}
\end{figure}

To probe the transient electronic structure upon below-gap pumping, TrARPES experiments (see schematic illustration in Fig.~2a) were performed with a tunable pump photon energy down to 160 meV and a fixed probe photon energy of 6.2 eV,  both generated by a laser amplifier running at a repetition rate of 10 kHz with a pulse width of 35 fs.  Figure 2b,c shows TrARPES snapshots measured  with pump photon energy of 160 meV and  polarization perpendicular to the scattering plane ($s$-$pol.$ pumping, where the pump polarization is along the armchair direction of black phosphorus). Floquet sidebands of the valence band (VB) are observed at $\Delta t$ = 0 ps in Fig.~2c  with an energy separation equal to the pump photon energy.

Moreover, the valence band maximum (VBM) shows a downward shift, which can be seen more clearly in the second-derivative images in Fig.~2d,e. To check if this is a rigid shift of the entire band or a momentum-dependent band renormalization, we extract the dispersions before and after pumping by analyzing the peak positions in the energy distribution curves (EDCs) shown in Fig.~2f.  The analysis shows that there is a remarkable  energy shift of 49 $\pm$ 3 meV at the $\Gamma$ point (bold curves), while away from the $\Gamma$ point, the band shift is largely reduced, and eventually becomes negligible for $|k|\geqslant0.08$~\textup{\AA}$^{-1}$.
Such momentum-dependent band renormalization clearly distinguishes from the surface photovoltaic effect, which induces a rigid shift of the entire VB due to a change in the surface potential \cite{Perfetti2D, Carpene2021, MonneySPVPRB}.
 The momentum-dependent band shift leads to a change of the effective mass from 0.10 $m_e$ to 0.15 $m_e$ in black phosphorus. 

In addition, the light-induced band renormalization upon below-gap pumping also exhibits a strong pump polarization dependence, with a strong renormalization when the pump pulse is polarized along the armchair (AC) direction (Fig.~S1 in the supplementary information \cite{supp}
), which is related to the coupling of light with the pseudospin degree of freedom \cite{Kim2020natmater,Zhou2023Nat}. We note that here optical absorption is largely suppressed as compared to above-gap excitation, and no signatures of multi-photon absorption or additional peaks below the absorption edge are observed \cite{oka2012nonlinear,li2022keldysh} (Fig.~S2 in the supplementary information \cite{supp}), suggesting that the light-induced change in the electronic structure upon 160 meV pumping  is not caused by excitons or other excitations, but rather by the light field, i.e. Floquet engineering.

\begin{figure*}[htbp]
	\centering
	\includegraphics[width=16.8 cm]{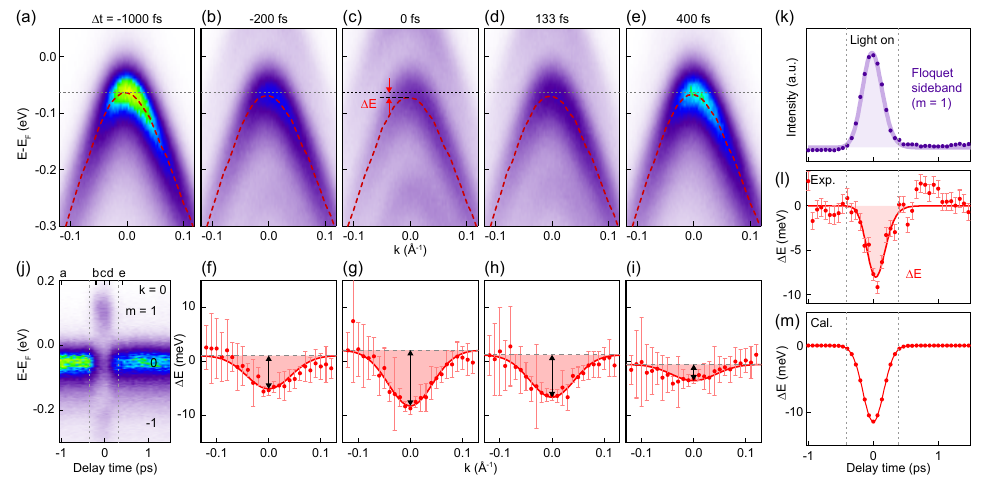}
	\caption{(a-e) TrARPES spectra at five different delay times. The extracted dispersions for the VB with $n$ = 0 are over-plotted as red dashed curves.  The data are measured along the direction at 30$^{\circ}$ from the AC direction with $p$-$pol.$ pump. The pump photon energy is 160 meV and the pump fluence is 317 $\mu$J/cm$^2$.  (f-i) Corresponding momentum-dependent energy shift.  (j) Intensity map of EDCs as a function of delay time.  (k) TrARPES intensity of $n$ = 1 sideband as a function of delay time.  (l) Energy shift at the $\Gamma$ point as a function of delay time. (m) Calculated energy shift at the $\Gamma$ point at different delay time.}
	\label{Fig3}
\end{figure*}

To further reveal the temporal evolution of the light-field-driven band renormalization, Figure 3a-e shows snapshots of the dispersion images measured at different delay times, where the extracted dispersions are over-plotted as dashed curves.
The momentum-dependent band shift at different delay times is extracted in Fig.~3f-i by subtracting the dispersion at $\Delta t$ = -1 ps, which shows a maximum value at $\Delta t$ = 0 (Fig.~3g) and decreases rapidly away from $\Delta t$ = 0 (Fig.~3h,i). Meanwhile, the intensity of the Floquet sidebands is also strongest at $\Delta t$ = 0, and these replica bands disappear for $\Delta t$ = 400 fs (Fig.~3e). To check whether the renormalization occurs simultaneously with the Floquet sidebands, Figure 3k,l shows a comparison of the temporal evolution of the Floquet sideband intensity (Fig.~3k) and the energy shift (Fig.~3l) extracted from the continuous evolution of EDC at the $\Gamma$ point (Fig.~3j). It is clear that the energy shift follows the same temporal evolution as the intensity of the Floquet sidebands, which is also reproduced by the Floquet tight-binding simulations (Fig.~3m). The simultaneous emergence of Floquet sidebands and momentum-dependent renormalization provides support for the realization of Floquet band engineering upon below-gap pumping.

To explore the evolution of Floquet engineering with pump photon energy, we show in Fig.~4a-e dispersion images measured at $\Delta t$ = 0 with pump photon energies tunable from 160 meV to 530 meV (the magnitude of the pump photon energy to the gap is schematically illustrated in Fig.~4f).
The extracted dispersions (red curves) are compared with those at $\Delta t$ = -1 ps (gray dashed curves) to visualize the light-induced renormalization.
The band renormalization is maximized at two pump photon energies: the lowest pump photon energy of 160 meV used in the measurements with the strongest Floquet interaction strength, and 325 meV which is near resonance with the band gap. For these two representative cases, the light-induced Floquet engineering shows different behavior: for below-gap pumping, the renormalization shows up as a flattening of the VBM; while for near-resonance pumping, there is a gap opening at the crossing point (marked as $\Delta$  in Fig.~1a respectively) and a renormalization ($\Delta$E) near the VBM.

\begin{figure*}[htbp]
	\centering
	\includegraphics[width=16.8cm]{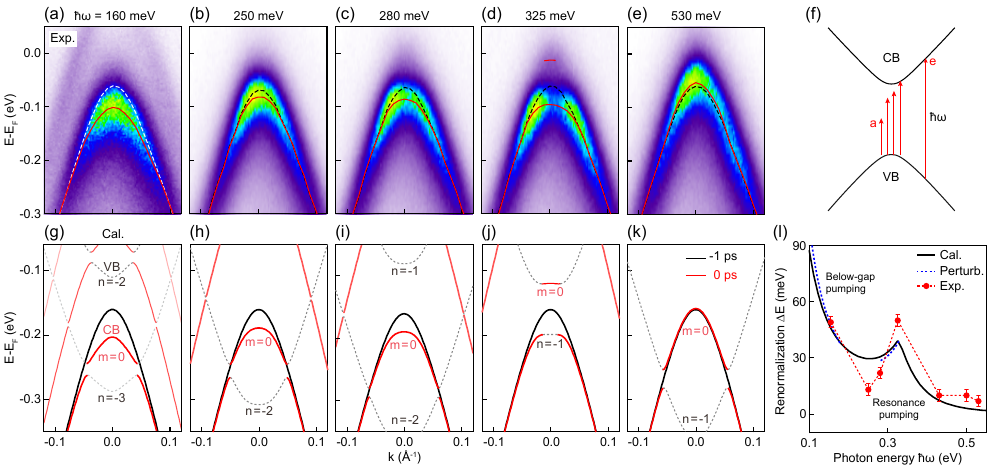}
	\caption{(a-e) TrARPES spectra with five different pump photon energies at the delay time of $\Delta t$ = 0 with pump fluence of 500 $\mu$J/cm$^2$. The red lines and gray dashed lines indicate the extracted dispersions at $\Delta t$ = 0 ps and corresponding data at $\Delta t$ = -1 ps. (f) A schematic of band structure with different pump photon energies used.  (g-k) Calculated dispersions based on Floquet tight-binding simulations with the photon energies in (a-e). Black curves show the VB without laser pumping, red curves stand for VB,  related sidebands and the gray dotted curves stand for sidebands of VB (n = -3, -2 for (g), n = -2 for (h-i) and n = -1 for (j,k))  (l) Extracted (red points) and calculated (black lines) energy shift at $\Gamma$ point with pump photon energy. The blue lines indicate the perturbative results as the pump photon energy approaches 0.16 eV and 0.33 eV. The error bar of the energy difference is defined from the curve fitting results.}
	\label{Fig4}
\end{figure*}

To understand the evolution of the light-field-driven engineering of  black phosphorus from below-gap to above-gap pumping, we perform the Floquet tight-binding calculations and systematically analyze the transient band renormalization (see more details in \cite{supp}). The calculated electronic structures are shown in Fig.~4g-k, in which the momentum-dependent band renormalization around the VBM is observed. The largest band shift occurs at the $\Gamma$ point. With the increase of the momentum, the band shift decreases quickly. Such effect can be understood from the perturbation theory and is strongly influenced by the momentum-dependent light-matter interaction and the intrinsic electronic structure of black phosphorus (see Fig.~S3 and S4 in the supplementary information \cite{supp}). With the change of pump photon energies, the observed band renormalization occurs for all cases. Two large energy shifts are found with pump photon energy of 160 meV and 325 meV (Fig.~4g,j), consistent with experimental results. Figure 4l plots the extracted energy shift at the $\Gamma$ point as a function of pump photon energy from experimental results (red symbols) and the Floquet tight-binding calculations (black curve). For above-gap pumping, the light-induced renormalization increases when decreasing the pump photon energy and reaches a local maximum when the pump photon energy is on resonance with the band gap. Further reducing the pump photon energy leads to a reduction of the light-induced renormalization, which eventually increases again when the pump photon energy is low enough.

More insights can be obtained by theoretically analyzing the evolution of the energy shift at the $\Gamma$ point based on the perturbation theory.
In the non-resonance region with below-gap pumping, as the photon energy approaches the band gap ($\hbar\omega\rightarrow E_g$), the energy shift $\Delta E$ at the $\Gamma$ point can be approximated by $\frac{1}{4}\left(\frac{\hbar\omega  (\hbar\omega-E_g)^2}{e\lvert\gamma_1\rvert E_0}+\frac{2 e\lvert\gamma_1\rvert  E_0}{\hbar\omega }+2 (\hbar\omega -E_g)\right)$, where $E_g$ is the band gap of black phosphorus in equilibrium, $\gamma_1$ is the optical matrix element, $e$ is the charge of electron, and $E_0$ is the electric field of the pumping laser. As a result, $\Delta E$ becomes larger when the photon energy approaches the band gap, resulting in a maximum value for $\Delta E$ at the resonant condition.  In this limit, the energy shift is almost proportional to electric field strength $E_0$ when the pump photon energy is fixed, namely $\Delta E \sim E_0$. 

On the other hand, when the photon energy is small ($\hbar\omega \ll  E_g$), $\Delta E$ at the $\Gamma$ point can be approximated by $\frac{3e^2\lvert\gamma _1\rvert^2 E_0^2}{8\hbar\omega E_g}\left(\frac{1}{\hbar\omega}+\frac{1}{E_g}\right)$, which increases monotonically when decreasing the pump photon energy $\hbar\omega$. As depicted in Fig.~4l, a notable increase in the energy shift $\Delta E$ at the $\Gamma$ point is observed when reducing the pump photon energy. In contrast to near-resonance pumping, once we fix the photon energy with a small value and treat the pump fluence as the perturbation, the energy shift scales linearly with the pump fluence $F$, namely $\Delta E \propto F\sim E_0^2$. The experimental pump fluence dependence shows overall consistent with the theoretical predicted linear dependence (Fig.~S5 in the supplementary information \cite{supp}). In the intermediate pump photon energy range ($160  ~\text{meV}<\hbar\omega<E_g$), $\Delta E$ has a local minimum.

Furthermore, we would like to note that the emergence of the Floquet state results from the competition between the pumping laser field and the relaxation of the excitations scattered by other quasi-particles, such as electrons and phonons, both coherently and de-coherently \cite{Sato2020jpb}. If the pump photon energy is too small, the period of the driving laser field ($T=2\pi/\omega$) will become comparable to or even longer than the scattering time $\tau$, then the Floquet effects are expected to be smeared out by the collisions \cite{GierzWSe2}. In our TrARPES measurements, the successful observation of Floquet engineering at the lowest photon energy of 160 meV, corresponding to $T \approx 26~\text{fs}$, suggests that the scattering time in black phosphorus is longer than this value. Thus, the ultrafast results shown in this work provide a benchmark for the scattering time in black phosphorus, and our estimations are consistent with results obtained from pump-probe optical measurements \cite{wang2015ultrafast,wang2016ultrafast}.

In summary, Floquet engineering is successfully realized in black phosphorus  in the below-gap pumping region, and a more comprehensive picture regarding the Floquet engineering of semiconductors is provided. 
First of all, the lack of direct optical absorption upon below-gap pumping in principle can efficiently suppress the dissipation channels and reduce the heating effect \cite{Lindner2020,wang2018theoretical,Dehghani2014prb,Dalessio2014prx}. Secondly, while near-resonance pumping could be an effective approach to enhance the Floquet engineering in small-gap semiconductor \cite{Zhou2023Nat}, below-gap pumping is also useful in achieving a large light-field-driven band renormalization once the Floquet states are formed. In particular, from the perturbation theory,   the energy shift of Floquet engineering at low pump photon energies exhibits a quadratic dependence on the peak electric field $E_0$ and is a monotonically decreasing as a function of the photon energy $\hbar\omega$, resulting in a pronounced enhancement of the band renormalization. Finally, it is worth noting that the resonance pumping condition is strongly material-dependent on the actual gap value, while in contrast, below-gap pumping has no such restriction and can even provide a stronger Floquet interaction strength, which may provide a promising way to extend Floquet engineering to a wider range of materials.

\begin{acknowledgments}
  This work is supported by the National Natural Science Foundation of China (Grant No.~12234011), National Key R\&D Program of China (Grant No.~2021YFA1400100), National Natural Science Foundation of China (Grants No. 92250305, No. 52388201, No. 11725418)
\end{acknowledgments}

%


\end{document}